\documentclass[12pt]{article}
\usepackage[]{epsfig}
\usepackage[centertags]{amsmath}
\usepackage{amsfonts}
\usepackage{amssymb}
\usepackage[english]{babel}
\usepackage{amsmath, amsthm, slashed,url}

\begin{document}

\title{\bf  Galilean symmetry in generalized abelian Schr\"{o}dinger-Higgs models with and without gauge field interaction}
\author{
Lucas Sourrouille
\\
{\normalsize \it Universidad Nacional Arturo Jauretche, 1888, 
}\\ {\normalsize\it Florencio Varela, Buenos Aires, Argentina}
\\
{\footnotesize  sourrou@df.uba.ar} } \maketitle
\maketitle

\abstract{We consider a generalization of nonrelativistic Schr\"{o}dinger-Higgs Lagrangian by introducing a nonstandard kinetic 
term. 
We show that this model is Galilean invariant, we construct the conserved charges associated to the symmetries and realize the 
algebra of the Galilean group. In addition, we study the model in the presence of a gauge field. We also 
show that the gauged model is Galilean invariant. Finally, we explore relations between twin models and their solutions.  }

\vspace{0.5cm}

{\bf pacs}: 
11.30.-j, 11.10.-z, 11.15.-q


\vspace{1cm}
\section{Introduction}

A new type of classical field theories has been intensively investigated during the several last years.
These theories, named k-field models, are usually endowed with nonstandard kinetic terms that change the dynamics
of the model under investigation. The $k$-field models have application in cosmology  \cite{APDM, APDM1, APDM11, APDM2, APDM4},
strong gravitational waves \cite{MV}, dark matter \cite{APL} and ghost condensates \cite{a1, a2, a3, a4, a5}
and others. In particular, an interesting issue concern to the study of topological structures, where topologically nontrivial 
configurations, named topological k-solutions, can exist \cite{SG}-\cite{SG1012}.

In the recent years, theories with nonstandard kinetic term, named $k$-field models, have received much attention. 
The $k$-field models are mainly in connection with effective cosmological models \cite{APDM, APDM1, APDM11, APDM2, APDM4} 
as well as the tachyon matter\cite{12} 
and the ghost condensates \cite{a1, a2, a3, a4, a5}. The strong gravitational waves \cite{MV} and dark matter \cite{APL}, are 
also examples of 
non-canonical fields in cosmology. Also, topological structure of these models was analyzed \cite{SG}-\cite{SG1012}, showing that 
the 
$k$-theories 
can support topological soliton solutions both in models of matter as in gauged models. 
\\[3mm]
In this paper we propose to study a nonrelativistic Higgs $k$-model. Here, the nonstandard kinetic terms are 
introduced by a function $\omega$, which depend on the Higgs field. In particular we show that $\omega(\rho)$,
where $\rho=\phi^\dagger \phi$, is Galilean invariant and we will 
construct the conserved charges associated with this invariance. We, also, show that the model realize the algebra of the 
Galilean group, if we choose a particular $\omega$, i.e. 
\begin{equation}
\omega(\rho)= \rho^n
\end{equation}
Finally, we analyze a nonrelativistic gauge model with nonstandard kinetic
term. In particular, we will 
concentrate on the Jackiw-Pi model \cite{JP, JP1} with nonstandard kinetic terms. We, also, 
show that this model is Galilean invariant, realizing the algebra of the group.

\section{The model and its symmetries}
\label{4v}

Let us start by considering the $(2+1)$-dimensional Schr\"{o}dinger model governed by the action,
\begin{equation}
S =\int d^{3}x\Big( i\phi^\dagger \partial_0 \phi -\frac{1}{2m}|\partial_i \phi|^2 + \lambda |\phi|^2 \Big)\;, 
\label{Ac1}
\end{equation}
Here, $\phi(x)$ is a complex scalar field and $\lambda$ is a strength coupling constant. 
Also,the metric tensor is  $g^{\mu \nu}=(1,-1,-1)$.
\\
It is well know that the model (\ref{Ac1}) presents Galilean invariance \cite{Ha}. This means that action 
(\ref{Ac1}) is invariant under time and space translation, rotations, Galilean boost and the $U(1)$ symmetry. More precisely, the 
Schr\"{o}dinger model remains invariant under the following symmetry transformations::
\begin{enumerate}
\item
The infinitesimal time-translation of the field
\begin{equation}
\delta \phi= a\partial_0 \phi\;,
\label{S1}
\end{equation}
where the Hamiltonian is the conserved charge associated to this symmetry 
\begin{equation}
H= \int d^2x \Big( \frac{1}{2m}|\partial_i \phi|^2 + \lambda |\phi|^2 \Big) 
\label{Ac2}
\end{equation}
\vspace{0.1cm}
\item
The infinitesimal translation of the field 
\begin{equation}
\delta \phi= a_i \partial_i \phi\;,
\label{S2}
\end{equation}
which leads to the conservation of linear momentum
\begin{equation}
P_i=\frac{i}{2} \int d^2 x \Big(\phi^\dagger \partial_i \phi - \partial_i \phi^\dagger \phi \Big) 
\end{equation}
\vspace{0.1cm}
\item
The infinitesimal field transformation due to a rotation 
\begin{equation}
\delta \phi= \theta {\bf r}\times {\bf \partial}\phi \;,
\label{7}
\end{equation}
being $\theta$ the rotation angle. Here the conserved charge obtained from the Noether theorem is angular momentum, 
\begin{eqnarray}
J= \int d^2x \Big( -\mathcal{P}_1 x_2 + \mathcal{P}_2 x_1 \Big)\;,
\end{eqnarray}
where
\begin{eqnarray}
\mathcal{P}_i = \frac{i}{2}\Big(\phi^\dagger \partial_i \phi - \partial_i \phi^\dagger \phi \Big)
\end{eqnarray}
\vspace{0.1cm}
\item
The infinitesimal field transformation due to Galilean boost
\begin{equation}
\delta \phi= i m v_i r_i \phi -t v_i\partial_i \phi
\label{}
\end{equation}
which leads to the conservation of the following charge
\begin{eqnarray}
&&G_i=\int d^2x \Big(\mathcal{P}_i t -m x_i \rho \Big)
\label{gi}
\\[3mm]
&&\rho=\phi^\dagger \phi
\end{eqnarray}
\vspace{0.1cm}
\item
The Galilean invariance is completed with the inclusion of $U(1)$ symmetry
\begin{equation}
\delta \phi = i\alpha \phi
\end{equation}
Here, the mass operator  $M=m\int d^2x \rho$ is the conserved charge associated to this transformation.
\vspace{0.1cm}
\end{enumerate}

The algebra of the Galilean group may be realized by using the Poisson brackets for functions of the matter fields, which are 
defined from the symplectic structure of the Lagrangian at fixed time to be
\begin{eqnarray}
\lbrace F,G
\rbrace_{PB}=i\int d^2x \left( \frac{\delta F}{\delta \phi^\dagger(r)}
\frac{\delta G}{\delta \phi(r)}- \frac{\delta F}{\delta \phi(r)}
\frac{\delta G}{\delta \phi^\dagger(r)}\right)
\
\label{poisson}
\end{eqnarray}
In the particular case in which $F=\phi$ and $G=\phi^\dagger$ we have 
\begin{eqnarray}
[\phi(x), \phi(x')^\dagger] = -i \delta^2(x- x')
\label{cr}
\end{eqnarray}
Using the Poisson bracket relations the above conserved charges can be shown to realize the algebra of the Galilean group
\begin{eqnarray}
&&[P_i, P_j]= [P_i, H]= [J, H]= [G_i, G_j]= 0  
\nonumber \\[3mm]
&&[J, P_i] =\epsilon^{ij}P_j
\nonumber \\[3mm]
&&[J, G_i] =\epsilon^{ij}G_j
\nonumber \\[3mm]
&&[P_i, G_j] = \delta^{ij} mN
\nonumber \\[3mm] 
&&[H, G_i]= P_i
\label{galalb}
\end{eqnarray}
In this section we are interested in exploring a generalization of the model (\ref{Ac1}). Following the same idea of the works 
cited in Ref.\cite{SG, SG0, SG2, SG9, SG10, SG101}, we modify the 
model (\ref{Ac1}) by changing both the canonical kinetic term of the scalar field and the potential term, so that the proposed 
model is described by the action  
\begin{equation}
S = \int d^{3}x \;\;\omega(\rho) \mathcal{L}_{NR}=\int d^{3}x \;\;\omega(\rho)\Big( i\phi^\dagger \partial_0 \phi 
-\frac{1}{2m}|\partial_i \phi|^2 + \lambda |\phi|^2 
\Big)=S_1 + S_2 +S_3\;, 
\label{Ac3}
\end{equation}
where
\begin{eqnarray}
&&S_1 = \int d^{3}x \;\;\omega(\rho) i\phi^\dagger \partial_0 \phi
\nonumber \\[3mm]
&&S_2 = -\int d^{3}x \;\;\omega(\rho)\frac{1}{2m}|\partial_i \phi|^2
\nonumber \\[3mm]
&&S_3 = \int d^{3}x \;\;\omega(\rho) \lambda |\phi|^2
\label{three}
\end{eqnarray}
Here, the function $\omega(\rho)$ is in principle an arbitrary dielectric function of the complex scalar field $\phi$,
and $\rho$ is related $\phi$ by 
\begin{equation}
 \rho = \phi^\dagger \phi\;,
\label{}
\end{equation}
where $n$ is a positive integer.
\\
In the next we will 
calculate the variation of the action (\ref{Ac3}) under time and space translation, angular rotation, Galilean boost and U(1) 
transformation. 
We begin to calculate the variation of the action (\ref{Ac3}) under time and space translation
\begin{equation}
\delta \phi= a\partial_0 \phi
\label{S11}
\end{equation}
\begin{equation}
\delta \phi= a_i \partial_i \phi
\label{S22}
\end{equation}
The variation (\ref{S11}) implies 
\begin{equation}
\delta \omega(\rho)= \frac{\delta \omega}{\delta \rho}\delta \rho= a\frac{\partial \omega}{\partial \rho} \partial_0 \rho = 
a\partial_0 \omega 
\label{v1}
\end{equation}
So that,
\begin{eqnarray}
\delta S_1 &=& \int d^{3}x \;\;\Big[i\delta\omega(\rho) \phi^\dagger
\partial_0 \phi + i\omega(\rho)\delta (\phi^\dagger \partial_0 \phi) \Big]= \int d^{3}x \;\;\Big[ia \partial_0 \omega(\rho) 
\phi^\dagger
\partial_0 \phi + i\omega(\rho)\delta (\phi^\dagger \partial_0 \phi) \Big]
\nonumber \\
&=&
\int d^{3}x \;\;\Big[ia \partial_0 \omega(\rho) 
\phi^\dagger
\partial_0 \phi + i\omega(\rho)(a \partial_0\phi^\dagger \partial_0 \phi +a \phi^\dagger \partial^2̣_0\phi) \Big]
\end{eqnarray}
By integration by parts, the last term of this integral, we have,
\begin{eqnarray}
\delta S_1 =0
\end{eqnarray}
The variation of $S_2$ under (\ref{S11}) leads to
\begin{eqnarray}
\delta S_2 =-\frac{a}{2m}\int d^{3}x \;\;\Big[(\partial_i \partial_0\phi^\dagger \partial_i \phi + \partial_i  
\phi^\dagger\partial_i\partial_0 \phi)\omega(\rho)+|\partial_i \phi|^2 \partial_0 \omega(\rho)\Big]
\end{eqnarray}
Then, integrating by parts the first term of this integral, we immediately arrive to   
\begin{eqnarray}
\delta S_2 =0
\end{eqnarray}
Finally, it easy to check that,
\begin{eqnarray}
\delta S_3 =\lambda \int d^{3}x \;\; [\delta \omega(\rho) \rho + \omega(\rho) 
\delta \rho] = a\lambda \int d^{3}x \;\; 
\partial_0[ \omega(\rho) \rho]= 0\;
\end{eqnarray}
where we have supposed the boundary condition
\begin{eqnarray}
\lim_{t,x \to \infty}\phi=0
\end{eqnarray}
Thus, the model is invariant under time translation. Space translation, involves 
\begin{equation}
\delta \omega(\rho)= a_i\partial_i\omega(\rho)
\label{v2}
\end{equation}
Then we have,
\begin{eqnarray}
\delta S_1 =\int d^{3}x \;\;\Big[ia_i \partial_i \omega(\rho)  
\phi^\dagger\partial_0 \phi- ia_i\omega(\rho)(\partial_i \phi^\dagger \partial_0 \phi +\phi^\dagger \partial_0 \partial_i \phi) 
\Big]
\end{eqnarray}
Integrating by parts the last term of this integral we get
\begin{eqnarray}
\delta S_1 =0
\end{eqnarray}
The variation with respect to $S_2$ is 
\begin{eqnarray}
\delta S_2 =-\frac{1}{2m}\int d^{3}x \;\;\Big[\Big(a_i \partial_i^2  
\phi^\dagger\partial_i \phi + a_i \partial_i \phi^\dagger \partial_i^2 \phi\Big)\omega(\rho)  + |\partial_i \phi|^2 a_i 
\partial_i \omega(\rho)\Big] 
\Big]
\end{eqnarray}
It can be easily seen that integrating by parts the first term, the variation becomes zero.
\\
For $S_3$, we have 
\begin{eqnarray}
\delta S_3 =\lambda \int d^{3}x \;\; [\delta \omega(\rho) \rho + \omega(\rho) \delta \rho] = a_i\lambda \int d^{3}x \;\; 
\partial_i[ \omega(\rho) \rho] = 0
\end{eqnarray}
The model is also invariant under rotations. Indeed, we have from (\ref{7})
\begin{eqnarray}
\delta \phi= x_1\partial_2 \phi -x_2 \partial_1 \phi
\label{r}
\end{eqnarray}
Such that,
\begin{eqnarray}
\delta \omega(\rho)= \frac{\partial_\omega}{\partial \rho} (x_1 \partial_2 \rho -x_2\partial_1 \rho)= x_1 \partial_2 
\omega(\rho) -x_2\partial_1\omega(\rho)
\label{v3}
\end{eqnarray}

\begin{eqnarray}
\delta S_1 =i \int d^{3}x&\Big[&[x_1 \partial_2 \omega(\rho) -x_2 \partial_1 \omega(\rho)]\phi^\dagger \partial_0 \phi + 
\omega(\rho)\partial_0 \phi[ x_1 \partial_2\phi^\dagger -x_2\partial_1 \phi^\dagger]
\nonumber \\
&+&\omega(\rho)\phi^\dagger \partial_0[x_1 \partial_2\phi -x_2 \partial_1 \phi]\Big]
\end{eqnarray}
Integrating by parts, in $x_1$ and $x_2$, the last term of this variation,  we easily arrive to
\begin{eqnarray}
\delta S_1 =0
\end{eqnarray}
Variation with respect to $S_2$, requires a bit more attention. By using (\ref{r}) and (\ref{v3}) we have
\begin{eqnarray}
\delta S_2 =-\frac{1}{2m}\int d^{3}x &\Big[&\partial_i ( x_1 \partial_2\phi^\dagger -x_2\partial_1 \phi^\dagger) \partial_i 
\phi \omega(\rho)+ \omega(\rho) \partial_i \phi^\dagger \partial_i (x_1 \partial_2\phi -x_2 \partial_1 \phi) 
\nonumber \\
&+&
|\partial_i\phi|^2 [x_1\partial_2 \omega(\rho) - x_2 \partial_1 \omega(\rho)] 
\Big]
\label{1.2}
\end{eqnarray}
Developing the first two terms of this integral and after some algebra we can check that $\delta S_2 =0$,
\\
The invariance under rotation of the model is completed by the variation of $S_3$,
\begin{eqnarray}
\delta S_3 =\lambda \int d^{3}x \delta[\omega(\rho) \rho]&=& \lambda\int d^{3}x \Big[x_1\partial_2 \omega(\rho) -x_2 \partial_1 
\omega(\rho)\Big] \rho + \omega(\rho)\Big[x_1 \partial_2 \rho - x_2 \partial_1 \rho\Big]
\nonumber \\
&=& \lambda\int d^{3}x \Big[\partial_2[\omega(\rho) \rho x_1] - \partial_1[\omega(\rho) \rho x_2]\Big] =0 
\end{eqnarray}
Let us concentrate on the Galilean boost, 
\begin{eqnarray}
\delta \phi = (im v_i x_i -tv_i \partial_i)\phi
\label{g1}
\end{eqnarray}
Under this transformation the variation of $\omega(\rho)$ is 
\begin{eqnarray}
\delta \omega(\rho) =\frac{\partial \omega}{\partial \rho}\delta \rho =-tv_i\partial_i \rho \frac{\partial \omega}{\partial \rho} 
= -tv_i\partial_i 
\omega(\rho)
\label{g2}
\end{eqnarray}
Thus, we have for $S_1$ the following variation,
\begin{eqnarray}
\delta S_1 = i \int d^{3}x \Big[-t v_i\partial_i \omega(\rho) \phi^\dagger \partial_0 \phi - \omega(\rho) 
tv_i\partial_i\phi^\dagger \partial_0 \phi - \omega(\rho) \phi^\dagger tv_i\partial_i\partial_0\phi\Big]
\end{eqnarray}
Integrating by parts the last term, it is easy to check $\delta S_1 = 0$
\\
The variation of $S_2$ may be evaluated by using (\ref{g1}) and (\ref{g2}), so that
\begin{eqnarray}
\delta S_2 = -\frac{1}{2m} \int d^{3}x \Big[-t v_i\partial_i\omega(\rho)|\partial_i \phi|^2 + (-tv_i\partial_i^2\phi^\dagger 
\partial_i\phi - tv_i\partial_i \phi^\dagger \partial_i^2 \phi) \omega(\rho)\Big]\;,
\end{eqnarray}
which vanish after integrating by parts the last term of this integral. 
\\
The $S_3$ is also invariant under Galilean boost. Indeed we have,
\begin{eqnarray}
\delta S_3 = \lambda \int d^{3}x \Big[-t v_i\rho \partial_i\omega(\rho) - tv_i \omega(\rho) \partial_i \phi^\dagger \phi + tv_i 
\omega(\rho) \phi^\dagger \partial_i \phi \Big]\;, 
\end{eqnarray}
where the last term may be integrated by parts, arriving to $\delta S_3 = 0$ 
\\
Finally, the $U(1)$ invariance of (\ref{Ac3}) is automatically satisfied, since $\omega(\rho)$ and $\mathcal{L}_{NR}$ are $U(1)$
invariant, and then 
\begin{eqnarray}
\delta S = \int d^{3}x \Big(\delta \omega(\rho) \mathcal{L}_{NR} + \omega(\rho) \delta \mathcal{L}_{NR}\Big) = 0
\end{eqnarray}

Using the Noether it is not difficult to obtain the conserved charges associated to the Galilean symmetries. In particular we 
arrive 
to the following quantities:
\begin{eqnarray}
H= \int d^{2}x\;\; j_0 dx^2 = \int d^{2}x\;\; i\phi^\dagger \partial_0 \phi \omega(\rho) -\mathcal{L} = \int d^{2}x\;\;
\Big(\frac{1}{2m}|\partial_i \phi|^2 + \lambda |\phi|^2 \Big) \omega(\rho)\;,
\label{Homega}
\end{eqnarray}
which is the Hamiltonian of the model (\ref{Ac3}). 
\\
\begin{eqnarray}
P_i  = \frac{i}{2}\int d^{2}x \Big[\Big(\phi^\dagger\partial_i 
\phi -\partial_i\phi^\dagger\phi\Big)\omega(\rho) -\rho \partial_i \omega(\rho)\Big]
\label{Pomega}
\end{eqnarray}
This is the conserved charge associated to space-translations, which differs from the usual nonrelativistic $P_i$ in the fact 
that here we have the function $\omega(\rho)$ multiplying the term $\phi^\dagger\partial_i 
\phi -\partial_i\phi^\dagger\phi$.
\\
\begin{eqnarray}
J =\int d^{2}x \Big(-\mathcal{P}_1 x_2 + \mathcal{P}_2 x_1
\Big)\;,
\label{Jomega}
\end{eqnarray}
which is the usual expression of the Angular momentum. 
\\
\begin{eqnarray}
G =\int d^{2}x \Big(-m 
x_i\rho + \mathcal{P}_i 
t \Big)\omega(\rho)
\label{Gomega}
\end{eqnarray}
which differs from (\ref{gi}) only on the factor $\omega(\rho)$. 
\\
\begin{eqnarray}
N = -\alpha\int d^{2}x\;\; \omega(\rho)\rho
\label{Nomega}
\end{eqnarray}
which is a generalization of the usual mass operator.

\section{The Galilean Algebra}
\label{2v}

In this section we shall study the algebra of the generators associate to the symmetry transformations studied in section 
(\ref{4v}). We have seen in section (\ref{4v}) that the algebra of the Galilean group is realized by the Poisson bracket 
(\ref{poisson}). Also, the Poisson bracket (\ref{poisson}) implies the commutation relation (\ref{cr}), which is the fundamental 
relation to construct the algebra (\ref{galalb}). The commutator (\ref{cr}), is the usual commutator between the fundamental 
field of the theory and its canonical conjugate, which is usually defined as
\begin{eqnarray}
\pi = \frac{\partial\mathcal{L}}{\partial(\partial_0\phi)}= i\phi^\dagger
\end{eqnarray}
So that,
\begin{eqnarray}
[\phi (x), \pi(x')] = \delta^2(x-x')
\end{eqnarray}
However, if we apply this commutation relation to construct the Galilean algebra of the model  (\ref{Ac3}), it is not difficult 
to see that we can not construct the Galilean algebra (\ref{galalb}). For instance we can check easily that,
\begin{eqnarray}
[ P_i, H] \not= 0
\end{eqnarray}
where $P_i$ and $H$ are given by the expressions (\ref{Homega}) and  (\ref{Pomega}).  
The problem lies in the fact that, here, $\pi$ is 
not $i\phi^\dagger$. Indeed,  
\begin{eqnarray} 
\pi = \frac{\partial\mathcal{L}}{\partial(\partial_0\phi)}= i\phi^\dagger \omega(\rho)
\end{eqnarray}
So that $\pi$ is a function of $\phi^\dagger$ and $\phi$ and therefore the definition  (\ref{poisson}) of the Poisson bracket does 
not apply. For this reason, we must redefine the theory in terms of new fundamental fields. In general, this is difficult due to 
the arbitrariness of the function $\omega(\rho)$. However, if we choose
\begin{eqnarray} 
\omega(\rho)= \rho^n \;,
\end{eqnarray}
where, $n$ is an arbitrary positive real number, we can rewrite the model (\ref{Ac3}) as follows 
\begin{eqnarray}
S&=&\int d^{3}x \Big(i\phi^\dagger \partial_0\phi -\frac{1}{2m}|\partial_i\phi|^2 +\lambda \rho \Big)\omega(\rho) =  \int d^{2}x 
\Big(i\phi^\dagger \partial_0\phi \rho^n -\frac{1}{2m}|\partial_i \phi|^2 \rho^n + \lambda\rho^{n+1}\Big)
\nonumber \\
&=&\int d^{3}x \Big(i(\phi^{n+1})^\dagger \phi^{n}\partial_0\phi - \frac{1}{2m}\phi^n \partial_i \phi 
(\phi^n)^\dagger \partial_i\phi^\dagger  + \lambda (\phi^{n+1})^\dagger \phi^{n+1}\Big)
\nonumber \\
&=&\int d^{3}x \Big(\frac{i}{n+1} (\phi^{n+1})^\dagger \partial_0\phi^{n+1} - \frac{1}{2m(n+1)^2} 
\partial_i(\phi^{n+1})\partial_i(\phi^{n+1})^\dagger + \lambda (\phi^{n+1})^\dagger \phi^{n+1}
\label{Ac5}
\end{eqnarray}
From (\ref{Ac5}), it is natural to define new fields, such that 
\begin{eqnarray}
\psi = \phi^{n+1}\;, \;\;\;
\;\;\;
\psi^\dagger = (\phi^\dagger)^{n+1}\;,
\label{55}
\end{eqnarray}
Thus, the action (\ref{Ac5}) is rewritten as
\begin{eqnarray}
S = \int d^{3}x \Big( \frac{i}{n+1} \psi^\dagger \partial_0\psi - \frac{1}{2m(n+1)^2} 
\partial_i(\psi)\partial_i(\psi)^\dagger + \lambda \psi^\dagger \psi \Big)
\label{}
\end{eqnarray}
Comparing this action with the nonrelativistic action (\ref{Ac1}), we see immediately that both are very similar. Then, the 
canonical conjugate field is 
\begin{eqnarray}
\pi = \frac{\partial\mathcal{L}}{\partial(\partial_0\psi)}= \frac{i}{n+1}\psi^\dagger\;,
\end{eqnarray}
and we can 
define the Poisson bracket, in terms of the new fields, following the definition (\ref{poisson}),
\begin{eqnarray}
\lbrace F,G
\rbrace_{PB}=i\int d^2x \left( \frac{\delta F}{\delta \psi^\dagger}
\frac{\delta G}{\delta \psi}- \frac{\delta F}{\delta \psi}
\frac{\delta G}{\delta \psi^\dagger}\right)
\
\end{eqnarray}
In particular, if $F=\psi$ and $G=\psi^\dagger$ we recover the usual commutation relation between the fundamental field and its 
canonical conjugate,
\begin{eqnarray}
\lbrace \psi,\psi^\dagger
\rbrace_{PB}=[\psi, \psi^\dagger] = i\int d^2x \Big(-\delta^2(x- x')\delta^2(x- x')\Big)
= -i \delta^2(x- x')
\end{eqnarray}
We can proceed in the same way as with the action and check that the conserved charges (\ref{Homega}), 
(\ref{Pomega}), (\ref{Jomega}), (\ref{Gomega}), (\ref{Nomega}), writing in terms of the fields $\psi$ and $\psi^\dagger$, are 
identical to the conserved charges of the model (\ref{Ac1}).
Thus, the conserved charges written in terms of $\psi$ and $\psi^\dagger$ as well as the commutation relation between $\psi$ and 
$\psi^\dagger$, lead us to similar context of the nonrelativistic case analyzed in section (\ref{4v}). Therefore, it is easy to 
understand that the generalized model (\ref{Ac3}), with $\omega(\rho) = \rho^n$ satisfies the algebra of the Galilean group 
expressed in (\ref{galalb}). 
  
\section{Gauged model}
Let us consider the model, in which Higgs field is coupled to a gauge field $A_\mu(x)$,
\begin{equation}
S = S_A + \int d^{3}x \;\;\omega(\rho) \mathcal{L}_{NR}= S_A + \int d^{3}x \;\;\omega(\rho)\Big( i\phi^\dagger D_0 \phi 
-\frac{1}{2m}|D_i \phi|^2 + \lambda |\phi|^4 
\Big)\;, 
\label{Ac6}
\end{equation}
where the covariant derivative is 
\begin{eqnarray}
D_{\mu}= \partial_{\mu} + ieA_{\mu}\;\;\;\;\;\;,(\mu =0,1,2)
\end{eqnarray}
and $S_A$ denote the dynamics of the gauge field. In particular we will assume that $S_A$ is a $2+1$ dimensional Chern-Simons 
action, given by,
\begin{eqnarray}
S_{cs}=
 \frac{\kappa}{4} \int d^3x \epsilon^{\mu \nu \alpha}A_\mu F_{\nu
 \alpha}= \kappa \int d^3x \left( A_0 F_{12} + A_2 \partial_0 A_1
\right)
\end{eqnarray}
In the same form that in the model (\ref{Ac3}), it is not difficult to see that, the model (\ref{Ac6}) is invariant under time 
and space translations, angular rotation, Galilean boost and $U(1)$ transformation. In addition if we choose, $\omega(\rho) = 
\rho^n$, the model (\ref{Ac6}) may be rewritten as    
\begin{eqnarray}
S &=& \int d^{3}x \Big( i(\phi^{n+1})^\dagger [\frac{1}{n+1}\partial_0 \phi^{n+1} + ieA_0\phi^{n+1}] 
\nonumber \\
&-&\frac{1}{2m}(\frac{1}{n+1}\partial_i(\phi^\dagger)^{n+1} -ieA_i (\phi^\dagger)^{n+1}) 
(\frac{1}{n+1}\partial_i\phi^{n+1} + ieA_i \phi^{n+1})
\nonumber \\
&+& \lambda \rho^{n+2}\Big) + S_{cs}
\end{eqnarray}
In terms of the fields $\psi$ and $\psi^\dagger$,
\begin{eqnarray}
S &=& \int d^{3}x \Big( i \psi^\dagger [\frac{1}{n+1}\partial_0 \psi + ieA_0\psi] 
\nonumber \\
&-&\frac{1}{2m}(\frac{1}{n+1}\partial_i\psi^\dagger -ieA_i \psi^\dagger) 
(\frac{1}{n+1}\partial_i\psi + ieA_i \psi)
+ \lambda (\psi^\dagger \psi)^2\Big) + S_{cs}
\end{eqnarray}
Let us, now, define the action $S^{'}$, such that $S^{'}= (n+1)S$,
\begin{eqnarray}
S^{'} = \int d^{3}x \Big( i \psi^\dagger D_0^{'}\psi
-\frac{1}{2m}|D_i^{'} \psi|^2 
+ \lambda_1 (\psi^\dagger \psi)^2\Big) + S_{cs}^{'}
\label{Ac7}
\end{eqnarray}
where, here, the covariant derivative is defined as
\begin{eqnarray}
D_{\mu}^{'}\psi = \partial_{\mu}\psi + ie_1 A_{\mu}\psi\;\;\;\;\;\;,(\mu =0,1,2)\;,
\end{eqnarray}
the $S_{cs}^{'}$ is
\begin{eqnarray}
S_{cs}^{'}=
 \frac{\kappa_1}{4} \int d^3x \epsilon^{\mu \nu \alpha}A_\mu F_{\nu
 \alpha}= \kappa_1 \int d^3x \left( A_0 F_{12} + A_2 \partial_0 A_1
\right)\;,
\end{eqnarray}
and the coupling constants $e_1$, $\kappa_1$ and $\lambda_1$ are
\begin{eqnarray}
e_1= e(n+1),\;\;\;\;\;\; \kappa_1=\kappa (n+1) ,\;\;\;\;\;\; \lambda_1 = \lambda (n+1)
\end{eqnarray}
Thus, the model (\ref{Ac6}) may be rewritten in terms of fields $\psi$ and $\psi^\dagger$ as
\begin{eqnarray}
S = \int d^{3}x \frac{1}{n+1}\Big( i \psi^\dagger D_0^{'}\psi -\frac{1}{2m}|D_i^{'} \psi|^2 
+ \lambda_1 (\psi^\dagger \psi)^2\Big) + S_{cs}
\label{Ac8}
\end{eqnarray}
This is the well know Jackiw-Pi model \cite{JP, JP1}, which is Galilean invariant and satisfies the algebra of the formula 
(\ref{poisson}) inherent to the Galilean group. So, the model (\ref{Ac6}) realize the Galilean algebra. 

\section{Twin models}

We can also modify the model (\ref{Ac1}) by introducing two different dielectric functions
\begin{equation}
S = \int d^{3}x \;\;\Big[\omega_1(\rho)\Big( i\phi^\dagger \partial_0 \phi 
-\frac{1}{2m}|\partial_i \phi|^2\Big) + \lambda \omega_2(\rho)|\phi|^2 \Big]
\label{Ac9}
\end{equation}
As the model (\ref{Ac3}), it is easy to check that (\ref{Ac9}) is also Galilean invariant. Indeed, we can rewrite (\ref{Ac9}) in 
three separate actions as in formula (\ref{three}) 
\begin{eqnarray}
&&S_1 = \int d^{3}x \;\;\omega_1(\rho) i\phi^\dagger \partial_0 \phi
\nonumber \\[3mm]
&&S_2 = -\int d^{3}x \;\;\omega_1(\rho)\frac{1}{2m}|\partial_i \phi|^2
\nonumber \\[3mm]
&&S_3 = \int d^{3}x \;\;\omega_2(\rho) \lambda |\phi|^2
\label{three1}
\end{eqnarray}
and as we showed in section (\ref{4v}) each of the three actions are Galilean invariant for an arbitrary dielectric function. So, 
the actions written in formula (\ref{three1}) are Galilean invariant for arbitrary $\omega_1$ and $\omega_2$.
\\
Again the model do not satisfies the Galilean algebra for an arbitrary $\omega_1$ and $\omega_2$. However, if we choose $\omega_1 
=\rho^n$ and $\omega_2 = \rho^h$, with $n$ and $h$ arbitrary positive real numbers, we can rewrite (\ref{Ac9}) as
\begin{equation}
S = \int d^{3}x \;\;\Big( \frac{i}{n+1}(\phi^\dagger)^{n+1} \partial_0 \phi^{n+1} 
-\frac{1}{2m(n+1)^2}|\partial_i \phi^{n+1}|^2 + \lambda |\phi|^{2(h+1)} \Big)
\label{Ac10}
\end{equation}
if we define $\psi$ as in (\ref{55}) we have,
\begin{equation}
\phi = \psi^{\frac{1}{n+1}}
\end{equation}
so that 
\begin{equation}
S = \int d^{3}x \;\;\Big( \frac{i}{n+1}\psi^\dagger \partial_0 \psi 
-\frac{1}{2m(n+1)^2}|\partial_i \phi^{n+1}|^2 + \lambda |\phi|^{2\frac{h+1}{n+1}} \Big)
\label{Ac11}
\end{equation}
Writing in this form it is evident that the model (\ref{Ac9}) satisfies the Galilean algebra. We can proceed in the same way for 
the gauged model. 
\\
Finally, let us concentrate on the solutions of the deformed model (\ref{Ac3}) . Here, we are interested on the 
static field configurations that minimize the energy functional associated to the model (\ref{Ac3}). Thus, for 
the model (\ref{Ac3}), we have
\begin{equation}
E= \int d^2x \Big( \frac{1}{2m}|\partial_i \phi|^2 + \lambda |\phi|^2 \Big) \omega(\rho) 
\label{Ec2}
\end{equation}
In the particular case that we choose the coupling constant to be
\begin{equation}
\lambda = \frac{1}{2m}
\end{equation}
the theory is governed by the Hamiltonian
\begin{equation}
E= \int d^2x \Big( |\partial_i \phi - \phi|^2 + \partial_i\rho \Big)\frac{\omega(\rho)}{2m} 
\label{Ec3}
\end{equation}
The last term of this expression may be written as a total derivative if 
\begin{equation}
\omega(\rho) = \frac{\partial f}{\partial \rho}
\label{}
\end{equation}
which, may be supposed without loss of generality. In this way we have,
\begin{equation}
E= \int d^2x \Big( |\partial_i \phi - \phi|^2 \frac{\omega(\rho)}{2m} + \partial_i f(\rho)\Big)
\label{Ec4}
\end{equation}
The total derivative may be dropped with the hypothesis that $f(\rho)$ is well-behaved. Then, the energy is bounded below by 
zero, and this lower bound is saturated by solutions to the first-order self-duality equation
\begin{equation}
\partial_i \phi= \phi
\label{7}
\end{equation}
The solution of this equation satisfies, not only, the Euler-Lagrange equation of (\ref{Ec2}), but also Euler-Lagrange equation 
of the model (\ref{Ac1}), i.e.
\begin{equation}
\partial_i^2 \phi= \phi
\end{equation}
Here, it is important to note that the Euler-Lagrange equation following from the functional (\ref{Ec2}) is different from that 
following from the model (\ref{Ac1}). Thus, we conclude that the solution of the equation (\ref{7}) solve the Euler-Lagrange 
equations of an infinitely large family of theories parametrized by the functional $\omega(\rho)$. 
This deformation procedure has been used recently by many authors \cite{bp1, bp2, bp3, bp4, bp5, bp6, bp7} to obtain relations 
between 
similar models and their solutions.

\vspace{0.3cm}
In summary we have proposed a generalization of the nonrelativistic Schr\"{o}dinger-Higgs model. We have shown that this 
generalized model admits Galilean invariance and we have also explored its twin models and their solutions. In addition we show 
the Galilean invariance of a generalization of the Jackiw-Pi 
model.

\vspace{0.6cm}
{\bf Acknowledgements}\\
I would like to thank Department of physics at Univesidad de Buenos Aires for hospitality.
This work is supported by CONICET.

\end{document}